\documentclass[showpacs,aps,prb,twocolumn,superscriptaddress,amsmath,amssymb]{revtex4}

\usepackage{graphicx}
\usepackage{dcolumn}
\usepackage{bm}
\begin{document}
\title{Quantum distortion caused by magneto-elastic coupling for antiferromagnetic tetrahedral cluster of spin-1/2}%
\author{Ikumi \surname{Honda}}
\altaffiliation[Present address:]{
Ceratech Japan Co., Ltd.\\
500 Shinonoi Okada, Nagano 381-2295, Japan
}
\affiliation{Graduate School of Science and Technology}
\author{Kiyosi Terao}
\email{terk005@shinshu-u.ac.jp}
\affiliation{Graduate School of Science and Technology}
\affiliation{Faculty of Science, Shinshu University, 1-1, Asahi-3, Matsumoto 390-8621, Japan}
\date{\today}
\begin{abstract}
We study the effects of magneto-elastic coupling on the degenerate ground spin-state of the antiferromagnetic Heisenberg model on a regular tetrahedron of spin-1/2.
When displacement of spin is considered as a classical variable,
the degeneracy of the spin-states is lifted through the distance dependence of exchange coupling,
i.e., a kind of Jahn-Teller effect takes place.
On the other hand, when displacement is considered as a quantum-mechanical variable,
the degeneracy of the ground spin-state is not lifted
although the tetrahedron is distorted by quantum fluctuation caused by magneto-elastic coupling for one component of the normal coordinates of the doubly degenerate mode. 
We propose a new model for the structural phase transition of vanadium and nickel spinels: the tetragonal distortion is caused by quantum fluctuation and there exists a kind of hidden order with respect to the non-magnetic ground spin-states.
\end{abstract}
\maketitle

\section{ \label{sec:intro} Introduction }
The highly degenerate antiferromagnetic ground state of the pyrochlore lattice has received considerable attention.~\cite{pwan}
It is interesting how the degeneracy is lifted to satisfy the Nernst-Planck theorem.~\cite{lutt}   
Reduction of the degeneracy due to lattice distortion through the distance dependence of exchange parameter was investigated before by Terao~\cite{tera, ter2} with classical spins on the basis of the helical spin structure theory.~\cite{yosh, naga}
Detailed investigations into the spin Jahn-Teller or Peierls effect were given with classical spin tetramer-model by Tchernyshyov {\it et al.}~\cite{tche, tch2}
and with spin-1/2 tetramer-model for vanadium-spinels by Yamashita and Ueda~\cite{YamashitaUeda00} in the context of the resonating valence bond approach.~\cite{pwan2}
In these works, however, the displacement of spin is represented by classical and static variables.
Recently, quantum-mechanical treatment of the displacement of spin has given by the present authors \cite{HondaTerao1} for a spin-1/2 regular-triangular cluster.
They have obtained distortion due to quantum fluctuation caused by the  magneto-elastic coupling without lifting the degeneracy,
which is a very different aspect from the Jahn-Teller mechanism through lifting the degeneracy of the ground spin-states obtained by the static treatment. 

In the present paper, we investigate the effects of the magneto-elastic coupling upon the degenerate ground spin-state of antiferromagnetic (AF) Heisenberg Hamiltonian for a regular tetrahedron of spin-1/2 by treating displacement of spin quantum-mechanically.
We propose a new model for the structural phase transition of vanadium and nickel spinels:~\cite{yueda, mamiya,crawford}
the tetragonal distortion is caused by quantum fluctuation and there exists a hidden spin order with respect to the doubly degenerate non-magnetic ground spin-states.

\section{\label{sec:spin} Spin states }
\begin{figure}[hbt]
\includegraphics[width=3.8cm]{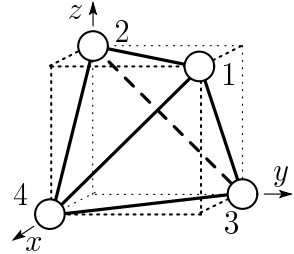}%
\caption{ Labels of spins and orientation of coordinates.}
\label{fig:fig01}
\end{figure}

\begin{figure}[hbt]
	\begin{center}
		\includegraphics[width=6.6cm]{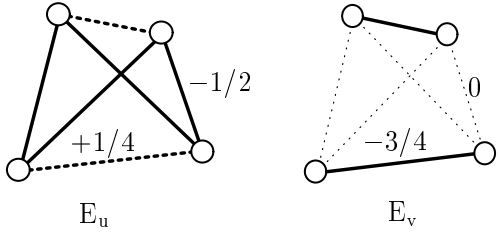}
	\end{center}
	\caption{$<\!\!{\rm E}\eta|\bm{s}_\ell \cdot \bm{s}_{\ell '}|{\rm E}\eta\!\!>$ for the E spin-states.}
			\label{fig:SE}
\end{figure}

The AF Heisenberg Hamiltonian for spins on a regular tetrahedron shown in Fig.~\ref{fig:fig01} is
\begin{align} 
	{\cal H}_0 = -2J_0 {\sum}_{\ell<\ell '} \bm{s}_\ell \cdot \bm{s}_{\ell '},
\end{align}
where $J_0 <$ 0 and $s$ = 1/2. The symmetry of ${\cal H}_0$ is of the T$_{\rm d}$ ($\bar{4}$3m) point group. We represent $2^4$ spin states by the $z$  components of $\bm{s}_\ell$ and classify them by total $S_z$ as shown in Table \ref{tab:SpinState}.
\begin{table}[hbt]
\caption{\label{tab:SpinState} Spin states and $S_z$.}
  \begin{ruledtabular}
    \begin{tabular}{lc}
    $S_z$ & State vectors\\
    \hline
     \ 2   & $|++++\!>$\\
     \ 1   & $|+++-\!>, \ |++-+\!>, \ |+-++\!>,\ |-+++\!>$\\
     \ 0   & $|++--\!>, \ |+-+-\!>, \ |+--+\!>,$\\
            & $|-++-\!>, \ |-+-+\!>, \ |-++-\!>$\\
     $-1$ & $|---+\!>, \ |--+-\!>, \ |-+--\!>,\ |+---\!>$\\
     $-2$ & $|----\!>$ 
    \end{tabular}
  \end{ruledtabular}
\end{table}
The $S_z=\pm 2$ states belong to the A$_1$ representation.
Each set of the quadruple $S_z=\pm 1$ states is reduced to  A$_1$ + T$_2$.
The hexaple $S_z=0$ states are reduced to A$_1$ + E +T$_2$.
We write the bases of the representation as $|\,{\rm A}_1,\, {\rm S}_z\!\!>$ for the singlet A$_1$, $|\,{\rm T}_2 \tau,\, {\rm S}_z\!\!>$ with $\tau$ = 1, 2, 3 for the triplet T$_2$, and $|\,{\rm E} \eta,\, S_z\!\!>$ with $\eta$ = u, v for the doublet E  representations. The eigen-energies are classified by the total spin $S$ as
\begin{subequations}
\begin{align}
	&\mathcal{H}_0 |\,{\rm A}_1,\,S_z \!\!>\  = -3J_0 |\,{\rm A} _1,S_z \!\!> &\text{for } S=2,\\
	&\mathcal{H}_0 |\,{\rm T}_2\tau,\,S_z \!\!>\  
      = J_0 |\,{\rm T}_2 \tau,S_z \!\!> &\text{for } S=1,\\
	&\mathcal{H}_0 |\,{\rm E} \eta \!\!>\  = \ 3J_0 |\,{\rm E} \eta \!\!> &\text{for } S=0.
\end{align}
\end{subequations}
For the AF case ($J_0<0$), the ground state belongs to the doublet E representation. The spin correlations ${<\!\!{\rm E}\eta|\bm{s}_\ell \cdot \bm{s}_{\ell '}|{\rm E}\eta\!\!>}$ for the E spin-states are shown in Fig.~\ref{fig:SE}.
\section{\label{sec:coup} Magneto-elastic coupling}
\begin{figure}[hbt]
	\begin{center}
		\includegraphics[width=7.2cm]{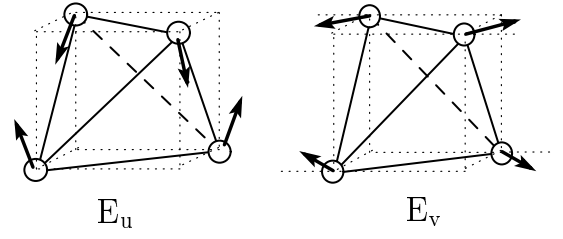}
	\end{center}
	\caption{The E-mode distortion.}
			\label{fig:QE}
\end{figure}

Assuming the exchange parameter $J$ depends on the distance between spins,
we investigate the effects of the magneto-elastic coupling on the E spin-state. 
We denote the small deviation of spin around the undistorted position $\bm{R}^{0}_{\ell}$ by $\bm{u}_\ell$.
Normal coordinates of spins are determined by reducing the twelve-dimensional representation by $\bm{u}_{\ell}$'s to A$_1$ + E + T$_1$ + 2\,T$_2$.
The distortions of the cluster are classified into A$_1$ ($Q_{\rm A}$, its normal coordinate), T$_2$ ($Q_1,\, Q_2,\, Q_3$) and E ($Q_{\mathrm{u}},\, Q_{\mathrm{v}}$) modes after eliminating the isotropic rotation T$_1$ and the uniform translation T$_2$.
The normal modes are written in the twelve-dimensional vector $ (\bm{u}_{1};\ \bm{u}_{2};\ \bm{u}_3;\ \bm{u}_4)$ as
 \begin{align}
	Q_{\mathrm{A}}
	(  1,  1,  1;  -1, -1,  1; -1,  1, -1; 1, -1, -1 )/{2\sqrt{3}}
\label{eq:QA}
\end{align}
for the singlet A$_1$ mode,
\begin{subequations}
	\begin{align}
		Q_1&( \, 0, \, 1, \, 1; \, 0, \, 1, -1; 
						\, 0, -1, \, 1; \, 0, -1, \, -1)/{2 \sqrt{2}},\\
		Q_2&( \, 1, \, 0, \, 1; \, 1, \, 0, -1;
						-1, \, 0, \, -1; -1,\, 0, \, 1)/{2 \sqrt{2}},\\
		Q_3&( \, 1, \, 1, \, 0; -1, -1, \ 0;
						1, -1, \, 0; -1, \, 1, \, 0)/{2 \sqrt{2}},
	\end{align}	\label{eq:QT2}%
\end{subequations}%
for the triplet T$_2$ mode, and
\begin{subequations}
	\begin{align}
	Q_{\rm u}&(1, 1, -2; -1, -1, -2;
		-1, 1, 2; \, 1, -1, 2)/{2 \sqrt{6}} ,\\
	Q_{\rm v} & (\, 1, -1,\, 0; -1, \, 1, \, 0;
		-1, -1, \, 0; \, 1, \, 1, \, 0)/{2 \sqrt{2}}
	\end{align}	\label{eq:QEvector}%
\end{subequations}%
for the doublet E mode.
The E mode distortion is illustrated in Fig. \ref{fig:QE}.
The bases of the irreducible representation made from bilinear combinations of the spin operators are 
\begin{align}
	f_{\rm A} =\sum_{\ell < \ell'}
					\bm{s}_\ell \cdot \bm{s}_{\ell '} /{\sqrt 6}
\label{eq:fA}
\end{align}
for the A$_1$ representation,
\begin{subequations}
\begin{align}
	f_1 = & \  (\bm{s}_1\cdot\bm{s}_4-\bm{s}_2\cdot\bm{s}_3)/{\sqrt{2}},\\
	f_2 = & \ (\bm{s}_1\cdot\bm{s}_3-\bm{s}_2\cdot\bm{s}_4)/{\sqrt{2}},\\
	f_2 = & \ (\bm{s}_1\cdot\bm{s}_2-\bm{s}_3\cdot\bm{s}_4)/{\sqrt{2}},		
\end{align}
			\label{eq:fT2}	%
\end{subequations}%
 for the T$_2$ representation, and
\begin{subequations}
	\begin{align}
	&f_{\rm u}=[(\bm{s}_1+\bm{s}_2)\cdot (\bm{s}_3+\bm{s}_4)\notag \\
	& \hspace{8em} -2(\bm{s}_1  \cdot \bm{s}_2 +\bm{s}_3 \cdot \bm{s}_4)] /{2\sqrt{3}},\\
	&f_{\rm v} = (\bm{s}_1 - \bm{s}_2) \cdot (\bm{s}_3 - \bm{s}_4)/2
	\end{align}
			\label{eq:fE}%
\end{subequations}%
for the E representation.~\cite{tche, tch2, YamashitaUeda00}
The perturbation Hamiltonian is composed of the normal vibrations and couplings of $Q_\alpha $'s with $f_\alpha$'s: 
\begin{align}
	{\mathcal{ H'}} =& \frac{1}{2m} ( {P_{\rm A}}^2 
	+{P_1}^2+{P_2}^2+{P_3}^2 + {P_{\rm u}}^2+{P_{\rm v}}^2) \notag \\
	 +& \frac{m}{2}  \bigl[ {\omega_{\rm A}}^2 {Q_{\rm{A}}}^2
	  +{\omega_{\rm T}}^2 ( {Q_1}^2 + {Q_2}^2+ {Q_3}^2) \notag \\
	 &\ \ \ \ \ \ \  +{\omega_{\rm E}}^2 ( {Q_{\rm u}}^2 + {Q_{\rm v}}^2) \bigr] \notag \\
	 -&2 \bigl[ J'_{\rm{A}}Q_{\rm{A}}f_{\rm {A}}
		+J'_{\rm{T}}(Q_1 f_1+ Q_2 f_2+ Q_3 f_3) \notag \\
	 &\ \ \ \ \ \ \  +J'_{\rm{E}}(Q_{\rm u} f_{\rm u}+ Q_{\rm v} f_{\rm v}) \bigr],
			\label{eq:perturbation1}
\end{align}
where $J'_\alpha $'s are coupling constants due to the change in $J$ by distortion. When $J$ depends only on the distance between spins, we have
\begin{equation} 
    J'_{\rm A} = 2 J', \ \  J'_{\rm T} = \sqrt{2} J', \ \  J'_{\rm E} = J'.
\end{equation}
where $J'=\partial J/\partial a$ and $a=| \bm{R}_\ell^0 -\bm{R}_{\ell'}^0 |$.

In the two dimensional subspace of the ground spin-states, $f_\alpha $'s are given in the 2$\times$2 matrix form as
\begin{subequations}
\begin{align}
	f_{\rm A} &= -\frac{\sqrt{6}}{4}\sigma_{\rm e}, \\
	f_{\rm u}  &= -\frac{\sqrt{3}}{2}\sigma_z, \ \ 
	f_{\rm v}  = -\frac{\sqrt{3}}{2}\sigma_x,\\
\text{\hspace{-10em} \, and }
 \ \ \ \ \ \ \ \ \ \ \ \ 
f_\tau  &= 0, \ \ \  \tau=1,2,3, \ \	\label{eq:fT_E}
\end{align}
\label{eq:f_inE}
\end{subequations}
\hspace{-8 pt} where $\sigma_{\rm e}$, $\sigma_z$ and $\sigma_x$ are the unit matrix  and the Pauli matrices.
Moreover, the product representation ${\rm T}_2 \times {\rm E}$ is reduced to ${\rm T}_1 + {\rm T}_2$, which does not contain the E representation, hence the perturbation belong to the T$_2$ representation is irrelevant in the subspace of the ground spin-states. 

\section{\label{sec:stat} Static theory }
First, we consider the effect of the static displacement of spin. Now, the perturbation is
\begin{align}
	\mathcal{H'} =&	\frac{ m }{2} \bigl[ {\omega_{\rm{A}}}^2{Q_{\rm{A}}}^2
	 +{\omega_{\rm{E}}}^2 ( {Q_{\rm u}}^2 + {Q_{\rm v}}^2) \big] \notag \\
	& -2 \big[ J'_{\rm{A}}Q_{\rm{A}}f_{\rm {A}} 
	  +J'_{\rm{E}}(Q_{\rm u} f_{\rm u}+ Q_{\rm v} f_{\rm v})\bigr].	
		\label{eq:staticH}
\end{align}
In the subspace of $|{\rm E}\eta \!\!>$'s,
\begin{align}
	\mathcal{H'} = \frac{ m}{2} &\big[ {\omega_{\rm{A}}}^2 {Q_{\rm{A}}}^2
	 +{\omega_{\rm{E}}}^2 ( {Q_{\rm u}}^2 + {Q_{\rm v}}^2) \big]
	+\frac{\sqrt3 J'_{\rm{A}}Q_{\rm{A}}}{\sqrt2} \notag \\
	&+\sqrt{3} \big[ 
		J'_{\rm{E}}(Q_{\rm u} \sigma_z + Q_{\rm v} \sigma_x ) \big],
			\label{eq:staticHp}
\end{align}
Then the eigenvalues are
\begin{align}
  \delta E' =\frac{m}{2} & \big[ {\omega_{\rm{A}}}^2 {Q_{\rm{A}}}^2
	 +{\omega_{\rm{E}}}^2 ( {Q_{\rm u}}^2 + {Q_{\rm v}}^2) \big]
  +\sqrt{\frac{3}{2}}J'_{\rm{A}}Q_{\rm{A}} \notag \\
   & \mp \sqrt{3}J'_{\rm{E}}\sqrt{{Q_{\rm u}}^2+{Q_{\rm v}}^2}.
\label{eq:staticE}
\end{align}
The degeneracy of the ground spin-state is lifted. Minimizing $\delta E'$, we have
\begin{subequations}
\begin{align}
&\delta E'_{\rm min} 
=- \frac{ 3 {J_{\rm A}'}^2} { 4m { \omega_{\rm A} }^2}
   - \frac{ 3{J_{\rm E}'}^2 }{2 m {\omega_{\rm E}}^2 },\\
\hspace{-3em} \text{at }\ &Q_{\rm A} = -\sqrt{\frac{3}{2}}
                \frac{ J'_{\rm A} }{ m {\omega_{\rm A}}^2 }, \ 
	\sqrt{{Q_{\rm u}}^2+{Q_{\rm v}}^2}
	=\frac{\sqrt{3} |J'_{\rm E}| }{ m {\omega_{\rm E}}^2 }.
			\label{eq:statQ}
\end{align}
\end{subequations}

The degeneracy of the ground spin-state is lifted by distortion through ${Q_{\rm u}}^2+{Q_{\rm v}}^2$ although the ratio of $Q_{\rm u}$ and $Q_{\rm v}$ is arbitrary up to quadratic order as shown by Yamashita and Ueda.~\cite{YamashitaUeda00}
The coefficients of the linear combination for the separated spin-state depend on the ratio of $Q_{\rm u}$ and $Q_{\rm v}$,
so the perturbed ground state is non-degenerate with respect to the spin state but not unique with respect to the shape of the cluster. 
Thus a kind of Jahn-Teller mechanism is obtained by the static model although the energy gain due to the distortion is determined by ${Q_{\rm u}}^2+{Q_{\rm v}}^2$.
These results are compared with the quantum mechanical results in the next section. 

\section{\label{sec:dyn} Dynamical theory }
Next, we consider the effect of dynamical displacement of spin.
By making use of the creation and annihilation operators, $b^\dagger_\alpha$ and $b_\alpha$, for normal mode $Q_\alpha $, 
we rewrite Eq.~(\ref{eq:perturbation1}) as
\begin{align}
  {\mathcal H'}  = \sum_\alpha \Bigl[
    &\hbar \omega _\alpha \bigl( b_\alpha ^\dagger b _\alpha +\frac{1}{2} \bigr) \notag \\
    &-\sqrt{ \frac{ \hbar }{ 2m\omega _\alpha } } 
     {J'} _\alpha \bigl( b_\alpha + b_\alpha ^\dagger \bigr) f_\alpha 
     \Bigr].
              \label{eq:perturbationBB}
\end{align}
Introducing modified operators
\begin{align}
  \tilde{b}_\alpha = b_\alpha - \frac{\sqrt2 J'_\alpha f_\alpha}{ \sqrt{m\hbar \omega_\alpha^3 } },
  \  \tilde{b}_\alpha^\dagger
    = b_\alpha^\dagger - \frac{\sqrt2 J'_\alpha f_\alpha}{ \sqrt{m\hbar \omega_\alpha^3 } } ,  
          \label{eq:modifiedBB}
\end{align}
we have
\begin{equation}
  \mathcal{H'} =\sum_{\alpha }
  \Big[
    \hbar\omega_\alpha
    \Big(
      \tilde{b}_\alpha^\dagger \tilde{b}_\alpha +\frac{1}{2} 
    \Big)
  -\frac{2}{m {\omega_\alpha }^2 } {{J'}_{\alpha }}^2 f_{\alpha}^2
  \Big].
             \label{eq:H1tild}
\end{equation}
Commutation relations of the modified operators are
\begin{equation}
[\tilde{b}_\alpha ,\tilde{b}_\alpha^\dagger ] =1, \ 
[\tilde b_\alpha ,\tilde b_\alpha ] =[\tilde b_\alpha^{\dagger} ,\tilde b_\alpha^{\dagger} ]=0,
\end{equation}
so $\tilde b_\alpha$, $\tilde b_\alpha^\dagger$ are the Boson operators.
Because the commutation relation of $f_{\rm u}$ and $f_{\rm v}$ is
\begin{align}
	[f_{\rm u}, f_{\rm v}] &=
	i \sqrt{3}/2 \{ ( \bf{s}_1 - \bf{s}_2) \cdot ( \bf{s}_3 \times \bf{s}_4) \nonumber\\
	&+( \bf{s}_3 - \bf{s}_4) \cdot ( \bf{s}_1 \times \bf{s}_2) \},
\end{align}
we have 
\begin{align}
	[ \tilde{b}_{\rm u}, \tilde{b}_{\rm v} ^\dagger ] = [\tilde{b}_{\rm u}, \tilde{b}_{\rm v}] 
	&= [\tilde{b}_{\rm u}^\dagger, \tilde{b}_{\rm v}^\dagger] \notag \\
	=  i \frac{\sqrt{3} {J'_{\rm E}}^2 }{ m \hbar \omega_{\rm E}^3} 
	\bigl\{[ & ( \bf{s}_1 - \bf{s}_2) \cdot ( \bf{s}_3 \times \bf{s}_4) \notag \\
	&+( \bf{s}_3 - \bf{s}_4) \cdot ( \bf{s}_1 \times \bf{s}_2) \bigr\},
\label{eq:comm_btilde}
\end{align}
which characterizes the chirality of the spin tetrahedron. 
Although the description of the excited states consistent with Eq.~(\ref{eq:comm_btilde}) are complicated,
the ground state with respect to $\tilde{b}_\alpha$ is simply defined as
\begin{align}
	\tilde{b}_\alpha  |\,\Gamma\gamma,S_z\!>_0 \,=0, 
\end{align}
which is equivalent with
\begin{equation}
  b_\alpha |\, \Gamma\gamma, S_z \!>_0 
  \,=\sqrt{ \frac{2}{m\hbar{\omega_\alpha}^3 } }
    J'_\alpha f_\alpha | \,\Gamma\gamma, S_z \!>_0 \!.
			\label{eq:TeigiSiki}
\end{equation}
 In the subspace of the ground spin-states $|\, {\rm E}\eta \!\!>_0$'s,
\begin{equation}
  \mathcal{H'} = \sum_{\alpha}
     \Bigl[ -\frac{ 2 }{ {m\omega_\alpha}^2 } {J'_\alpha}^2 
       { f_{\alpha} }^2 + \frac{ \hbar\omega_\alpha }{ 2 } \Bigr],
			\label{eq:HpE}
\end{equation}
where ${f_{\alpha} }^2$ 's are proportional to $\sigma_{\rm e}$, the $2 \times 2$ unit  matrix, by Eqs. (\ref{eq:f_inE}).
Then, the change in energy is
\begin{equation}
  \delta E' = -\frac{3}{ 4m\omega_{\rm A}^2 }{ J'_{\rm{A} }}^2
  - \frac{ 3 {J'_{\rm E}}^2 }{ m \omega_{\rm{E}}^2 }.
			\label{eq:dy_dEp}
\end{equation}
Note that the degeneracy is not lifted although the energy is decreased by distortions.
The contribution of the A$_1$ mode to $\delta E'$ obtained here is equal with that by the static model, and the contribution of the E mode is twice that by the static model. 

In the subspace of $|\, {\rm E}\eta \!\!>_0$'s,
\begin{align}
  & Q_{\rm A} =-\sqrt{\frac{3}{2}} \frac{J'_{\rm A}}{ m{\omega _{\rm A}}^2 } \sigma_{\rm e},\\
  & Q_{\rm u} =-\frac{\sqrt{3}J'_{\rm E}}{ m {\omega_{\rm E} }^2 } \sigma_z,
   \ \ Q_{\rm v}  = -\frac{\sqrt{3} J'_{\rm E}}{ m {\omega_{\rm E} }^2 } \sigma_x.
			\label{eq:QE}
\end{align}
The distortion due to the A$_1$ mode in $|\, {\rm E}\eta \!\!>_0$ is
\begin{align}
	<\! {\rm E}\eta| Q_{\rm A} | {\rm E}\eta \!\!>_0  = -\sqrt{\frac{3}{2}} \frac{J'_{\rm A}}{ m{\omega _{\rm A}}^2 }
\end{align}
for $\eta$ = u and v, and those due to the E$_{\rm u}$ mode for $|\, {\rm Eu} \!\!>_0$ and $|\, {\rm Ev} \!\!>_0$ states are
\begin{subequations}
\begin{align}
 	<\! {\rm Eu}| Q_{\rm u} | {\rm Eu}\!\!>_0 =-&\frac{\sqrt{3}J'_{\rm E}}{ m {\omega_{\rm E} }^2 },\\
	<\! {\rm Ev}| Q_{\rm u} | {\rm Ev}\!\!>_0 =+&\frac{\sqrt{3}J'_{\rm E}}{ m {\omega_{\rm E} }^2 },
\end{align}
		\label{eq:Qu}
\end{subequations}
\hspace{-9pt} which have opposite sign according to the spin-states $|\, {\rm E}\eta\!\!>_0$ with $\eta$ = u or v.  In contrast to $Q_{\rm A}$ and $Q_{\rm u}$, $Q_{\rm v}$ is non-diagonal and its expectation values are vanishing:
\begin{align}
  <\! {\rm E}\eta| Q_{\rm v} | {\rm E}\eta \!\!>_0 =0,
	\label{eq:Qv}
\end{align}
i.e., the distortion due to E$_{\rm v}$ mode is smeared out and can not be observed.

The fluctuation is estimated as
\begin{align}
  <\! {\rm E}\eta | {Q_{\rm v}} ^2 | {\rm E}\eta \!\!>_0 
  =\frac{ 3{ J'_{\rm E} }^2 }{m^2 {\omega_{\rm E}}^4 }  +\frac{\hbar}{2m\omega_{\rm E}}
		\label{eq:Qv2}
\end{align}
for both $\eta$ = u and v. The term with $\hbar$ comes from the zero-point motion, and
the first term is equal to ${<\!{\rm E}\eta | {Q_{\rm u}} |{\rm E}\eta \!\!>_0^2}$ by Eq.~(\ref{eq:Qu}).
Although the tetrahedron distorts into different shapes depending on the spin-states $|\, {\rm E}\eta\!\!>_0$ with $\eta$ = u or v, the spin-states remain doubly degenerate. 

The expectation value of the squared displacement of the E mode distortion is
\begin{equation}
  <\!{\rm E}\eta | {Q_{\rm u}}^2+ {Q_{\rm v}}^2 |{\rm E}\eta \!\!>_0 
   =\frac{ 6{ J'_{\rm E} }^2 }{m^2 {\omega_{\rm E}}^4 }  +\frac{\hbar}{m\omega_{\rm E}}.
		\label{eq:kanaE} 
\end{equation}
On the other hand, the sum of squared expectation values is
\begin{align}
	<\!{\rm E}\eta | {Q_{\rm u}} |{\rm E}\eta \!\!>_0^2
	+ <\!{\rm E}\eta |{Q_{\rm v}} |{\rm E}\eta \!\!>_0 ^2
	=  \frac{ 3{ J'_{\rm E} }^2 }{m^2 {\omega_{\rm E}}^4 },
\end{align}
which is half of the first term in Eq.~(\ref{eq:kanaE}) apart from the contribution of the zero-point motion and equal to the value obtained by the static model, Eq.~(\ref{eq:statQ}). 
The change in energy due to the E mode distortion obtained by dynamical model is twice that obtained by static model because of the quantum fluctuation.

 Expanding $J$ up to the quadratic terms of $Q_\alpha$'s, we obtain the coupling with $f_{\rm A}$ for the A$_1$ representation as
\begin{align}
  \mathcal{H''}[{\rm A}_1] = &-\bigl[  J_{\rm AA}''{Q_{\rm A}}^2
        +\frac{ J_{\rm AE}'' }{ \sqrt2 } ({Q_{\rm u}}^2 +{Q_{\rm v}}^2 ) \notag \\
     & +\frac{ J_{\rm AT}'' }{ \sqrt3 } ({Q_1}^2+{Q_2}^2 +{Q_3}^2)        
    \bigl] f_{\rm A},
			\label{eq:H''A}
\end{align}
with $f_1$, $f_2$ and $f_3$ for the T$_2$ representation as
\begin{align}
  \mathcal{H''}[{\rm T}_2] = &
  -\sqrt2 J_{\rm TA}'' \bigl(
    Q_{\rm A} Q_1 f_1 + Q_{\rm A} Q_2 f_2+Q_{\rm A} Q_3 f_3
  \bigr) \notag \\
  - \frac{ J_{\rm TE}'' }{ 2\sqrt2 }
   \bigl\{& \bigl[ {Q_{\rm u}} Q_1 + Q_1 {Q_{\rm u}}
   -\sqrt3 \bigl( Q_{\rm v} Q_1+ Q_1 Q_{\rm v} \bigr) \bigr] f_1 \notag \\
   +& \bigl[ {Q_{\rm u}} Q_2 + Q_2 {Q_{\rm u}}
   +\sqrt3 \bigl( Q_{\rm v} Q_2+ Q_2 Q_{\rm v} \bigr) \bigr] f_2 \notag \\ 
      +&2\bigl( Q_{\rm u} Q_3 +Q_3 Q_{\rm u} \bigr) f_3 \bigr\} \notag \\
   - \frac{J''_{\rm TT}}{\sqrt2} 
    \bigl[ &\bigl( Q_2 Q_3+Q_3 Q_2 \bigr) f_1+\bigl( Q_3 Q_1+Q_1 Q_3 \bigr) f_2 \notag\\
      &+\bigl( Q_1 Q_2+Q_2 Q_1 \bigr) f_3 \bigr],
			\label{eq:H''T}
\end{align}
and with $f_{\rm u}$ and $f_{\rm v}$ for the E representation as
\begin{align}
  \mathcal{H''}[{\rm E}] =& -\sqrt2 J_{\rm EA}''
   \left(
     Q_{\rm A} Q_{\rm u} f_{\rm u}+Q_{\rm A} Q_{\rm v} f_{\rm v}
   \right) \notag \\
  - \frac{ J_{\rm EE}'' }{ \sqrt2 }&
   \bigl[({Q_{\rm u}}^2 - {Q_{\rm v}}^2) f_{\rm u}
      +(Q_{\rm u} Q_{\rm v} +Q_{\rm v} Q_{\rm u}) f_{\rm v} \bigl] \notag \\
   - J''_{\rm ET}& 
    \bigl( \frac{ Q_1^2+Q_2^2-2Q_3^2 }{ \sqrt6 } f_{\rm u}
      +\frac{ {Q_1}^2-{Q_2}^2 }{ \sqrt2 } f_{\rm v} \bigr).
			\label{eq:H''E}
\end{align}
When $J$ is a function of only the distance between spins,
\begin{align}
  &J_{\rm AA}''= {2\sqrt6 J''}/3,\, J_{\rm AE}'' = {J''}/{\sqrt3}, \notag \\
  &J_{\rm AT}'' = {3}/{\sqrt2}\Bigl( {J'}/{a}+{2J''}/{3}\Bigr), \notag \\
  &J_{\rm EA}'' = -2/\sqrt{3}J'',\, J_{\rm EE}''=-{J''}/{\sqrt6},\, J_{\rm ET}''=\sqrt2 J'',
\end{align}
where $J''=\partial^2 J/\partial a^2$.

In $\mathcal{H''}[{\rm A_1}]$, ${Q_{\rm u}}^2 + {Q_{\rm v}}^2$ is proportional to $\sigma_{\rm e}$ by Eq.~(\ref{eq:QE}) and ${Q_\tau}^2$'s vanish apart from the zero-point motion by Eqs.~(\ref{eq:fT_E}) and (\ref{eq:TeigiSiki}).
In $\mathcal{H''}[{\rm E}]$, ${Q_{\rm u}}^2 - {Q_{\rm v}}^2$ and $Q_{\rm u} Q_{\rm v} +Q_{\rm v} Q_{\rm u}$ vanish by Eq.~(\ref{eq:QE}). The last term vanishes by Eqs.~(\ref{eq:fT_E}) and (\ref{eq:TeigiSiki}).
Thus, $\mathcal{H''}$\ is proportional to $\sigma_{\rm e}$ in the subspace of $|E\eta, S_z\!>_0$\,'s.
Note that if $Q_\eta$ is considered classical, $Q_{\rm u} Q_{\rm v} +Q_{\rm v} Q_{\rm u}$ does not vanish,
so the degeneracy is lifted by $f_{\rm v}$.
After straightforward calculations, we obtain
\begin{align}
&\delta E'' =  \Bigl[
		\frac{3\sqrt6}{8} \frac{{J'_{\rm A} }^2} { m^2 {\omega_{\rm A}}^4 }
		+\frac {\sqrt3 \hbar} {4m \omega_{\rm A}}
	\Bigr] J''_{\rm AA}\notag 
	+ \frac{3\sqrt2\hbar}{8m\omega_{\rm T}}J''_{\rm AT} \notag \\
    &+\Bigl[
		\frac{3\sqrt3}{2} \frac{ {J'_{\rm E} }^2 } { m^2 {\omega_{\rm E}}^4 } +\frac{\sqrt3 \hbar} {4 m \omega_{\rm E}}
	\Bigr] J''_{\rm AE} 
	 + 
		\frac{3\sqrt3}{2} \frac { {J'_{\rm A}}^2 {J'_{\rm E}}^2 J''_{\rm EA}} 
                                   { (m \omega_{\rm A}\omega_{\rm E})^4 }.
\label{eq:kanaMA}
\end{align}

Now let us consider about a new mechanism for the tetragonal distortion at the phase transition without AF ordering in vanadium and nickel spinels with spin-1.~\cite{yueda, mamiya, crawford}
Yamashita and Ueda~\cite{YamashitaUeda00} have considered the mechanism by breaking up the tetrahedron of spin-1 composing  pyrochlore structure into the tetramer of spin-1/2 and by taking into account the spin-lattice coupling by the static model. They have explained the mechanism as a result of splitting of the degeneracy of the ground spin-state due to distortion of the cluster by considering higher terms more than quadratic terms to settle the ratio of $Q_{\rm u}$ and $Q_{\rm u}$.

 On the basis of the present dynamical model, the distortion of $Q_{\rm u}$ component emerges because the distortion of $Q_{\rm v}$ component is smeared out by the quantum-mechanical fluctuation.   
According to our model, the tetragonal distortion at the structural phase transition implies the  occurrence of  hidden ordering of the nonmagnetic spin-state $| {\rm E}\eta \!\!>_0$, $\eta$ = u or v, 
because the sign of  $<{\rm E}\eta | Q_{\rm u}| {\rm E}\eta \!\!>_0 $ for $\eta$ = u or v is opposite to each other.

\section{\label{sec:con} Discussion and conclusions}
We have studied the frustrating quantum spin-1/2 system on the regular tetrahedron by taking into account the effect of distortion up to the quadratic terms.
When the distortion is represented by the classical variables, the degeneracy of the ground spin-state is lifted by  the distance dependence of exchange parameter. The energy of separated ground spin-state is determined through ${Q_{\rm u}}^2+{Q_{\rm v}}^2$, hence the ratio of $Q_{\rm u}$ and $Q_{\rm v}$ is left arbitrary, which gives  continuous degeneracy with respect to the shape of the tetrahedron for the separated spin-state.

When the distortion is represented by the quantum-mechanical variables, the degeneracy of the modified spin-states $| \rm Eu\!\!>_0$ and $| \rm Ev\!\!>_0$ is not lifted by the magneto-elastic coupling although the cluster is distorted.
The change in energy due to the distortion obtained by dynamical model is twice that obtained by the static model. 

The distortions by $Q_{\rm A}$ (A$_1$ mode) and $Q_{\rm u}$ (E mode) do not fluctuate and their expectation values are equal to the results obtained by the static model apart from the contribution of the zero point motion.
On the other hand, $Q_{\rm v}$ (E mode) is smeared out by  the quantum fluctuation and does not contribute to the deformation in appearance.
Then the shape of the distorted tetrahedron is determined by $Q_{\rm u}$.
The distortion of the tetrahedron is attributed to the quantum fluctuation caused by magneto-elastic coupling.

We have proposed a new model for the structural phase transition without AF ordering in vanadium and nickel spinels.~\cite{yueda, mamiya,crawford}
The tetragonal distortion is caused by quantum fluctuation in contrast to the spin Jahn-Teller effect by Yamashita and Ueda.~\cite{YamashitaUeda00} 
According to our model, the tetragonal distortion at the structural phase transition implies the  occurrence of  hidden ordering of the nonmagnetic spin-states $| {\rm Eu}\!\!>_0$ or $| {\rm Ev}\!\!>_0$.

\end{document}